\documentstyle[twocolumn,prl,aps]{revtex}
\large

\begin{document}
\draft
\twocolumn[
\hsize\textwidth\columnwidth\hsize\csname @twocolumnfalse\endcsname

\draft
\title{
       Potential energy topology and relaxation processes in a model glass
      }
\author{
	F.~Demichelis$^{1}$  
	G.~Viliani$^{1}$,
	G.~Ruocco$^{2}$,
       }
\address{
	 $^{1}$ 
	 Universit\'a di Trento and Istituto Nazionale di Fisica
	 della Materia, I-38050, Povo, Trento, Italy.\\
	 $^{2}$ 
	 Universit\'a di L'Aquila and Istituto Nazionale di Fisica 
	 della Materia, I-67100, L'Aquila, Italy. 
	}
\date{\today}
\maketitle
\begin{abstract}
%============= abstract ================
We use computer simulation to investigate the topology of the potential 
energy $V(\{{\bf R}\})$ and to search for doublewell potential's (DWP)
in a model glass . By a sequence of Newtonian and dissipative dynamics we
find different minima of $V(\{{\bf R}\})$ and the energy profile along the 
least action paths joining them. At variance with previous suggestions, we 
find that the parameters describing the DWP's are correlated among each 
others. Moreover, the trajectory of the system in the 3$N$-d configurational
phase space follows a quasi-1-d manifold.  The motion parallel to the path
is characterized by jumps between minima, and is nearly uncorrelated from
the orthogonal, harmonic, dynamics.
%=======================================
\end{abstract}
]
%=============== text ==================
Among many unanswered questions concerning the dynamics of topologically 
disordered materials, two are related to the low energy excitations.
The first one regards the excitations responsible for the low temperature
thermal properties of glasses. Since the pioneering works of Phillips 
\cite{phil} and Anderson et al. \cite{ande}, who introduced the Two Level 
Systems (TLS) as the origin of the anomalous specific heat behaviour below 
1 K, many works were devoted to identifying the degrees of freedom associated 
to the TLS in real glasses, but a definite answer has not yet been reached. 
The second question regards the excitations responsible for: ({\it i}) the 
thermal behavior in the 1$\div$10 K region, where a plateau in the thermal
conductivity and a peak in $c_p/T^3$ are observed \cite{elliott}; and 
({\it ii}) the boson peak appearing in the inelastic neutron scattering
(INS)\cite{BPNDAP,BZNGP,WPCF,MNPSZ} and Raman spectra
\cite{CCFMR,KSPQ,BFBCT,GLYX,MNPSZ}. This second question is also largely 
unanswered, and different hypotheses on the nature of these excitations have
been proposed, ranging from an excess of acoustic-like modes to highly
anharmonic and localized excitations. A link between the above questions 
is suggested by the so called Soft Potential Model (SPM) \cite{spm}, which 
assumes the existence of few higly anharmonic degrees of freedom with a 
potential energy function described by a fourth order polynomial. 
The coefficients of the polynomial are supposed to be uncorrelated, 
and to give rise to a large variety of curves, including doublewell 
potentials (DWP's). Among the DWP's, those that are isolated,  
have low asymmetry and have barrier height resulting in a tunneling
splitting around 1 K, are candidate to be TLS.

Computer Simulations (CS) of model glasses might help to answer the
previous questions, and, in particular, to verify the existence of TLS
and to clarify the nature of DWP in glasses. To our knowledge the direct 
CS inspection of the passage from minimum to minimum in glasses is limited 
to a few cases.  Among them we mention the search for TLS performed by 
Stillinger and Weber \cite{WS85} and by Heuer and Silbey \cite{HS93,HS}.  
The conclusions of these works can be summarized as follows:
{\it i)} low barrier bistable degrees of freedom exist in binary glasses
with considerable variation in their physical parameters (barrier height, 
asimmetry, minima distance, etc..); 
{\it ii)} these parameters are strongly correlated, DWP
with large asymmetry typically have large barrier heights; 
{\it iii)} in agreement with SPM, the coefficients of the fourth order
polynomial describing the potential energy of the reaction coordinate, are
found to be uncorrelated;
{\it iv)} using a factorized probability distribution function for these
coefficients, it is possible to try an estimate of the number of TLS
per atom;
{\it v)} typically from 5 to 10 atoms participate in the motion in the DPW.

In this letter we investigate by CS the topology of the potential energy 
hypersurface of a Lennard-Jones glass with the aim to check whether 
the TLS can be observed in this system, the characteristics of the DWP's,
and the extent to which the harmonicity of the system is affected by the 
presence of DWP's. 
We will discuss the procedures used to search the minima of $V(\{{\bf R}\})$
and to identify the Reaction Coordinate (RC) and the Least Action Path (LAP)
joining pairs of minima. The main conclusions of the present work are the
following: 
{\it a)} no isolated double-wells are found in the system, i.e. in the 
investigated temperature range ($T > 5$ K) all the 
minima pertain to a network, and the thermally activated jumps among them 
are not controlled only by the energetic barriers, because, depending on
temperature, a significant role is also played by the entropic term \cite{R96}.
{\it b)} At variance with previous findings \cite{HS93,HS} (see item {\it
iii)} above) and with the hypothesis at the base of the SPM \cite{spm}
the coefficients of the polynomial representation of the energy profile felt
by the RC, are found to be higly correlated. {\it c)} The degrees of freedom 
orthogonal to the LAP feel an almost harmonic  potential, whose curvature is 
independent on the RC, i.~.e. on the position along  the LAP itself.

We investigated a sample of $N=864$ atoms interacting via the 6-12 Lennard-Jones 
potential, here we use $\epsilon/K_B=125.2$ K and  $\sigma=0.3405$ nm, appropriate 
for Argon. A microcanonical molecular dynamics simulation \cite{MD} is carried out 
at the fixed density of 42 mol/dm$^3$, and at different temperatures $T$ ($T=$ 15, 
12, 10, 7.5 and 6 K). Each run consists of a sequence of newtonian and dissipative 
dynamics. The newtonian trajectory is followed for 1 ps (50 integration time steps), 
subsequently a modified steepest descent method procedure (relaxation-like dynamics) 
is applied to quench the system and to find the inherent configuration $\{{\bf X}_i\}$
($i=1...N$), corresponding to a local minimum. The newtonian dynamics is then 
restarted from the same point in phase space where it was interrupted before
the  quenching, and the procedure is repeated  up to 25,000 times for 
each temperature. The two minimum configurations obtained with successive quenching, 
$\{{\bf X}^a_i\}$ and $\{{\bf X}^b_i\}$, are often the same.

The adopted procedure ensures that different minima, when found,
are close to each other in the 3$N$-d configurational space. About 130  different
minima have been found in the present work: their energies are 
spread in an interval of about 1 J/mole centered at about 300 J/mole above
the absolute  minimum representing the crystal. All the minima belong to an
inter-connected network and the system is observed to perform temporarily 
closed loop in the configurational phase space. 
In order to study the topopogy of the potential energy, and to find the easiest 
way to jump among neighbourING minima, a fast and efficient algorithm is developed 
to evaluate the  Least Action Path (LAP).  The LAP is determined as the path between 
${\bf X}^a_i$  and ${\bf X}^b_i$ that minimize the classical action integral,
$ \int_{\bf X^a}^{\bf X^b} ds \sqrt{ V({\bf R}(s)) - V_0} $, 
with $V_0= min \{ V(\{{\bf X}^a_i\}),V(\{{\bf X}^b_i\}) \}$. We studied 
$\approx$ 100 pairs of minima, evaluating the LAP joning them; in Fig.~1a we compare 
the potential energy profile evaluated along the LAP with that along the straight path.  
Although the LAP is not very far from being straight in the configurational space, 
as shown in Fig.~1b, the potential energy along the LAP is significantly 
lower, making the LAP itself highly preferred for the jump among minima. 
Moreover, the present analysis shows that the derivation of the DWP
properties from the profile of the potential energy along the straight 
path between minima gives unreliable results. The potential energy profile 
along the LAP between the minima $a$ and $b$ has been characterized by the 
energy difference between the two minima $\Delta V^m_{ab}$, their euclidean 
distance $D_{ab}$, the displacement of the atom that moves most $d_{ab}=max_i 
\{\vert {\bf X}^a_i-{\bf X}^b_i \vert \}$, the barrier height $\Delta V^B_{ab}$ 
(measured with respect to the higher minimum) and finally by the number of 
participation, defined as 
$P_{ab} = ( \sum_{i} \vert {\bf X}^a_i-{\bf X}^b_i \vert^2 )^2 
/ \sum_{i} \vert {\bf X}^a_i-{\bf X}^b_i \vert^4$. 
It turns out that: {\it i)} the participation is always around 20;
{\it ii)} among the 20 atoms involved in the jump, 1 or 2 account for 
90\% of the entire distance; {\it iii)} for each pair of minima $\Delta V^m$
is always higher than $\Delta V^B$.

Next we analyse the statistical correlation among the DWP's parameters.
In Fig.~2a we see that the different pairs of parametres that characterized 
the DWP are strongly correlated.
To give a statistical significance to the correlation, for each pair of 
parameters we calculated the correlation coefficient 
$r$ \cite{lcc}.
The values of $r$, reported in the first part of Table I, indicate statistical 
correlation between the parameters (with $n_{DWP}=141$, at the 5\% of significance 
level, the threshold for correlation is $r=0.165$). Two other representations of the
DWP are reported in the literature, these are the "$u$" representation, used in
the soft potential model \cite{spm}, 
$E=\epsilon \; u_0 \; [ u_2 (x/\sigma)^2 - u_3 (x/\sigma)^3 + (x/\sigma)^4 ]$,
and the "$w$" representation introduced in ref.~\cite{HS},
$E=\epsilon \; [ w_2 (x/\sigma)^2 - w_3 (x/\sigma)^3 + w_4 (x/\sigma)^4 ]$.
The SPM assumes that there is no correlation among the parameters of the set
$\{ u_0, u_2, u_3 \}$, while the authors of  ref.~\cite{HS}, noticing the correlation 
of the $u$ set, made use of the uncorrelation among $\{ w_2, w_3, w_4 \}$ to determine 
the number of TLS in model glasses. As can be seen in Figs.~2b and 2c, and in Table I, 
we found statistical correlation also in the case of the $w$ set. It is beyond the scope 
of the present work to explain this inconsistency, that can lie on the different glass 
examined or, more likely, on the different procedure used to determine the potential 
energy profile (LAP here, straight path in \cite{HS}).

In order to describe the low temperature anomalies of glasses, the most
important quantity is the total splitting of the ground state associated
to the TLS. To our knowledge, only one attempt has been made to estimate the
energy splitting due to the tunneling through TLS energy barriers. 
It {\it a)} assumes the existence of isolated pairs of minima and 
{\it b)} treats the problem as it were 1-dimensional (1-d) \cite{HS93}. 
From the present simulation, hypothesis {\it a)} can be neither cibfirmed nor
rejected because we did not directly identify any candidate TLS (i.~e. pairs
of minima with asymmetry $< 1$ K); recent results on Argon clusters
\cite{giorgia} indicate that pairs of minima with low asymmetry are "isolated"
in the sense that a third quasi-degenerate, adjacent minimum has not been observed.
As for item {\it b)}, the $(3N-3)$-dimensional tunneling problem can be
treated as an 1-dimensional one if the Schr\"odinger equation can be
factorized into $3N-3$ independent equations \cite{schiff}. We shall assume that
the reduction to $1-$D can be made under the less restrictive condition that
the relevant classical path (i.~e. the LAP) is independent of all the others.
To this end we have studied the curvatures of the potential energy 
surface along the LAP by calculating the dynamical matrix eigenvalues 
$\lambda_j(n)$ ($j=1,..,3N-3$) and the corresponding eigenvectors
in 42 ($n=1..42$) equally spaced configurations along the LAP itself.
In Fig.~3, we report, for each configuration along the LAP joining a typical
pair of minima, the 20 ($j=1..20$) lowest frequencies 
$\omega_j=\sqrt{\vert \lambda_j \vert}$ (we have assigned the minus sign to 
the frequencies associated to negative eigenvalues).
As can be seen, in the present case only one eigenvalue becomes negative, 
indicating a first order saddle point in the 3N-dimensional space, 
and only the lowest frequencies change appreciably along the LAP. 
In Fig.~4 we report, for the seven lowest eigenvalues, the projections of 
the eigenvectors on the local tangent at the LAP. 
Along the path, the main contribution comes from the lowest eigenvalue, 
while approaching the minima, an increasing contribution comes from
other eigenvalues. The same results are found for all pairs of minima. 
They indicate that the dynamics parallel and orthogonal to the reaction 
coordinate are nearly independent: quasi-harmonic vibrations control the 
orthogonal dynamics, by a set of eigenfrequencies that are independent of
the position along the path. The reaction coordinate is associated with 
a single eigenvalue (often the lowest one), as shown by the large value
of the projection of only one eigenvector in the path direction.

In conclusion, the picture emerging from the depicted scenario is that of a 
network of connected minima, where each pair is joined together by a 
1-dimensional path. The reaction coordinate along the path follows a
LAP in the 3N-configurational space, described by DWP with statistically
correlated parameters (asymmetry, height, etc.). The potential energy
experienced by the system along the LAP is significantly lower than 
that one along other paths, like, for example, the straight path. 
Orthogonally to the RC, the dynamics is harmonic, and the set of 
frequencies are nearly independent on the specific value of the RC. 
In the present simulation a single interconnected quasi-1-dim path has 
been found, indicating that no more than one DWP is active at the time. 
Extrapolating this result, we can state that the number of DWP's coexisting 
with {\it harmonic} excitations is $N_{DWP} < 10^{-3}$ atoms$^{-1}$.

We are grateful to acknowledge technical support by R.~Iori (CISCA,
Universit\'a di Trento), by V.~Dimartino (CASPUR, Roma) and by S.~Cozzini and 
X.~Voli (CINECA, Bologna).

\begin{table}
\caption
{The upper right part of the three matrices reports the
correlation coefficients $r$ among all the couple of parameters used to
describe the DWP in the physical representation $\{ \Delta V^m$, $\Delta
V^B$,
$D$, $d$, $P \}$, and in the two different polynomial representations
$\{ w_2$, $w_3$, $w_4\}$ and $\{ u_0$, $u_2$, $u_3\}$. The lower left part of
the correlation matrices indicates whether the two parameters are
($\bullet$)or are not ($\circ$) correlated at the 5\% level of confidence.}

\begin{tabular}{ccccccccc} 
			 &&           &              &              &        &       &&\\
	     \hfill\vline&& $\Delta V^m$ & $\Delta V^B$ & $D$       & $d$    & $P$   &&\\
\hline
	     \hfill\vline&&           &              &              &        &       &&\\
$\Delta V^m$ \hfill\vline&&     -     &     0.45     & 0.86         & 0.78   & 0.13  &&\\
$\Delta V^B$ \hfill\vline&& $\bullet$ &  -           & 0.72         & 0.72   & 0.02  &&\\
$D$          \hfill\vline&& $\bullet$ &  $\bullet$   &  -           & 0.85   & 0.25  &&\\
$d$          \hfill\vline&& $\bullet$ &  $\bullet$   &  $\bullet$   & -      & -0.15 &&\\
$P$          \hfill\vline&& $\circ$   &  $\circ$     &  $\bullet$   & $\circ$& -     &&\\
			 &&           &              &              &        &       &&\\
\hfill\vline&   $w_2$   &   $w_3$ & $w_4$&&      \hfill\vline&  $u_0$ & $u_2$   & $u_3$ \\
\cline{1-4} \cline{6-9}
\hfill\vline&           &         &      &&      \hfill\vline&      &        &       \\
$w_2$ \hfill\vline& -         & 0.49    & 0.17 &&$u_0$ \hfill\vline& - & -0.23   & -0.22 \\
$w_3$ \hfill\vline& $\bullet$ &  -      & 0.90 &&$u_2$ \hfill\vline& $\bullet$ &  -      &  0.87 \\
$w_4$ \hfill\vline& $\circ$   &$\bullet$&  -   &&$u_3$ \hfill\vline& $\bullet$ &$\bullet$&  -    \\
		  &           &         &      &&                  &         &       
\end{tabular}
\end{table}

{\footnotesize{
\begin{center}
{\bf CAPTIONS}
\end{center}

\begin{description}

\item  {Fig. 1 - 
a) Comparison between the potential energy profile along the LAP 
($\bullet$) and along the straight path ($\circ$) joining two typical
minima. $E_x$ is the energy of the crystalline minima.
b) Orthogonal distance between the LAP and the straight path as a function
of the reaction coordinate.
}

\item  {Fig. 2 - 
Example of correlation between the parameters describing the DWP. 
Each point represent the values found for the couple of parameters 
a) ($\Delta V^m$, $D$), b) ($w_4$, $w_3$), and c) ($u_3$, $u_2$) 
in the description of each DWP.
}

\item  {Fig. 3 - 
The 20 lowest eigen-frequencies $\omega=\sqrt{ | \lambda | }$ of the system 
evaluated in 42 equispaced atomic configurations along the LAP are reported
as a function of the reaction coordinate for a typical pair of minima. Negative
eigenvalues $\lambda$ are reported as negative eigen-frequencies $\omega$.
}

\item  {Fig. 4 - 
The projection on the LAP of the six lowest frequency eigen-vectors of the modes 
of Fig.~3 are reported as a function of the reaction coordinate.
}

\end{description}
}}
\end{document}